\documentclass[a4paper,final,12pt,table]{article}
\usepackage{amsmath,dbPaper2021}
\usepackage[round]{natbib}
\usepackage{SectionFonts}

\newcommand{\figs}{1}
\newcommand{\appndix}{0}
\newcommand{\submission}{0}
\newcommand{\figtabsname}{_figstabs.tex}
\newcommand{\appndixname}{_appendix.tex}
\newcommand{\abstrctname}{_abstract.tex}
\newcommand{\CurVersion}{\today}
\graphicspath{{../graphics/}{graphics/}}
\newcommand{\InitVersion}{October 13, 2020}
\AtBeginDocument{\let\oldref\ref\renewcommand{\ref}[1]{(\oldref{#1})}}
\renewcommand{\Authorspace}{\vspace{-1.5mm}}
\renewcommand{\Titlefont}{\fnt[22][21]}

\newcommand{\spce}{1.534}
\renewcommand{\MYTITLE}{On a Standard Method for Measuring the \\ Natural Rate of Interest}
\input{tcilatex}
\begin{document}

\title{\MYTITLE\thanks{%
Without implications, I am grateful to Adrian Pagan, Paolo Giordani, Claus
Brand, Jim Stock, Jesper Lind\'{e} and Kurt Mitman for comments. I thank two
anonymous referees for remarks that helped to improve previous versions of
the paper. I am grateful to the Jan Wallander and Tom Hedelius Foundation
and the Tore Browaldh Foundation for research support.} \\
\vspace{05mm}\setcounter{footnote}{0}}
\author{\href{http://www.danielbuncic.com}{{Daniel Buncic}}%
\setcounter{footnote}{2}\thanks{%
Corresponding author: Stockholm Business School, Stockholm University,
SE-103 37, Stockholm, Sweden. Email: \emailto{daniel.buncic@sbs.su.se}. Web: %
\url{http://www.danielbuncic.com}.} 
\affiliation }
\date{\DateVersion }
\maketitle

\AddAbstract\AddSecondTitlepage[0][0]

\setstretch{\spce}

\section{Introduction \label{sec:intro}}

\citeauthor*{holston.etal:2017}' (\citeyear{holston.etal:2017}) (simply
HLW's henceforth) estimates of the natural rate of interest have become a
benchmark or reference rate --- not only for policy makers at central banks
--- but also for asset managers and finance professionals needing to make
long-term investment decisions for their clients.

One of the reasons for the model's popularity and widespread use is its
simplicity; the core of the structural model is reminiscent of a New
Keynesian Policy model, albeit without a central bank reaction function (see
\cite{gali:2015} and \cite{buncic.melecky:2008}). Another reason is the
availability of computer code that is published on
the website of the \href{https://www.newyorkfed.org/research/policy/rstar}{%
Federal Reserve Bank of New York (FRBNY)}, which makes the non-standard
estimation of the model accessible to a broader audience. The fact that the
model is used to inform US policy makers about monetary policy decisions adds
further to the credibility and relevance of the model.\footnote{The model can
also be viewed as multivariate version of the unobserved component model of
\cite{clark:1987} (adding an inflation equation and allowing for interactions
between inflation and the output gap equations), or as a multivariate filter
(see \cite{benes.etal:2010}).}

In this paper, I\ show that HLW's implementation of \cites{stock.watson:1998}
(SW's) Median Unbiased Estimation (MUE)\ to determine the size of the
signal-to-noise ratio parameter $\lambda _{z}$ cannot recover the ratio of
interest $\frac{a_{r}\sigma _{z}}{\sigma _{\tilde{y}}}$ from MUE. This
inability to recover the ratio of interest is due to a misspecification in
HLW's Stage 2 model formulation. I show further that the structural break
regressions which are used as an auxiliary model in MUE are modified from
SW's original implementation in the simulations used to construct the
look-up values in Table 3 of their paper. The misspecification in the Stage
2 model --- together with the modification of the structural break
regressions --- leads to spuriously large estimates of the signal-to-noise
ratio parameter $\lambda _{z}$, which affects the severity of the downward
trend in other factor $z_{t}$, and thereby the estimates of the natural rate
of interest in HLW's model.

I provide a correction to the specification of the Stage 2 model that is
consistent with the signal-to-noise ratio $\lambda _{z}=\frac{a_{r}\sigma
_{z}}{\sigma _{\tilde{y}}}$ used in the estimation of the full structural
model of the natural rate, and further implement the structural break
regressions in line with the simulations employed by SW to construct the MUE\
look-up values. The correction that I provide is quantitatively important.
For the Euro Area, the UK, and Canada, the (corrected) MUE point estimates of
$\lambda _{z}$ are exactly $0$. The downward trend in the estimates of other
factor $z_{t}$ disappears entirely, with the $z_{t}$ estimates at (or very
close to) zero. For the US, the $\lambda _{z}$ point estimate shrinks from
$0.040$ to $0.013$, resulting in a much more subdued downward trend in other
factor $z_{t}$. The ensuing natural rate estimates are affected most strongly
at the end of the sample period. For the US, the (corrected)\ estimate of
$r_{t}^{\ast }$ is over $100$ basis points larger at approximately $1.5$
percentage points than from HLW's (misspecified) implementation of $0.48$.
For the Euro Area, $r_{t}^{\ast }$ is approximately $80$ basis points larger
at $1.03$ percentage points, instead of $0.24$, while for the UK\ and Canada,
the differences are more subtle, being respectively $45$ ($1.80$ percentage
points instead of $1.35$) and $27$ basis points larger ($1.73$ percentage
points instead of $1.46$).\ Estimates of trend growth remain unaffected by
the proposed correction to the Stage 2 MUE\ implementation.

This paper is related to --- but distinct from --- a broader literature on
estimating the natural rate of interest. For instance, \cite%
{berger.kempa:2019} use a Bayesian estimation approach to avoid pile-up at
zero problems, implementing the non-centered state-space parameterisation of
\cite{fruehwirth-schnatter.wagner:2010} for other factor $z_{t}$ to be able
to test if $\sigma _{z}$ is greater than zero.\footnote{\cite%
{berger.kempa:2019} also allow the variances of the shocks to the output gap
and trend growth equations to follow integrated stochastic volatility
processes, adding further a (latent) real rate cycle equation to HLW's
model, so the model is a somewhat modified version of HLW's original set-up.}
They find the posterior density of $\sigma _{z}$ to be centred at zero,
suggesting no variation in other factor $z_{t}$. \cite%
{lewis.vazquez-grande:2018} also employ a Bayesian approach to estimate a
less restrictive version of HLW's model, ie., one that specifies trend
growth and other factor $z_{t}$ to follow AR(1) processes instead of random
walks. They find that $z_{t}$ should be modelled as an AR(1) process with
transitory shocks, instead of following a random walk. \cite{kiley.etal:2020}%
, who also uses a Bayesian estimation approach, but on a model that does not
separate the natural rate into the sum of other factor $z_{t}$ and trend
growth, finds that there is little information in the data to determine the
process that generated the natural rate. These studies offer interesting
additional results on other factor $z_{t}$ and the natural rate. The
objective of the current paper is, nonetheless, to provide a correction to
HLW's Stage 2 model and MUE implementation, and to make it accessible to
users of this model.\footnote{Accompanying Matlab and R code files that
replicate the results presented here --- together with filtered and smoothed
estimates of the natural rate of interest, trend growth, other factor
$z_{t}$, and the output gap for all four countries --- are available from:
\url{http://www.danielbuncic.com}.}

As a final remark to readers familiar with (or interested in) the original
model of \cite{laubach.williams:2003} which first proposed MUE\ for the
estimation of the natural rate from a structural model, the problems that I
outline here with HLW's Stage 2 model and the MUE\ implementation are
exactly the same. This can be easily verified by expanding the matrices in
Section 4.4 \textquotedblleft \textit{The Stage 2 Model}\textquotedblright\
on page 4 of the PDF\ file: \textquotedblleft \textit{LW\_Code\_Guide.pdf%
\textquotedblright } contained in the replication code zip file posted on
the FRBNY website at %
\url{https://www.newyorkfed.org/medialibrary/media/research/economists/williams/data/LW_replication.zip}%
. The $\xi _{t}$ vector in equation (8) only contains one lag of trend
growth ($g_{t-1}$) and the extra $a_{5}$ parameter in the $\mathbf{H}%
^{\prime }$ matrix, resulting in a Stage 2 output gap equation that is
equivalent to HLW's misspecified formulation shown in the right column block
of \ref{cycl} below. The structural break regressions needed for MUE in Stage
2 are implemented in the same way as in HLW (see lines 109 to 114 in the R
file \texttt{rstar.stage2.R} which prepares the $y$ and $x$ data to be called
in the \texttt{median.unbiased.estimator.stage2.R} function implementing the
structural break regressions in lines 10 to 15).

The remainder of the paper is organized as follows. \Sref{sec:S2} gives a
brief outline of \HLW structural model of the natural rate of interest, and
contrasts how Median Unbiased Estimation\ is implemented in SW and in HLW. %
\Sref{sec:results} provides the empirical results. In \Sref{sec:concl}, the
study is concluded.

\section{Holston \textit{et al.}'s (2017) structural model and MUE\label%
{sec:S2}}

\subsection{Model set-up}

\HLW structural model of the natural rate takes the following form:\vsp[-2]%
\bsq\label{eq:hlw}
\begin{alignat}{3}
\text{Output}& \text{:} & & \hsp[23] & y_{t}& =y_{t}^{\ast }+\tilde{y}_{t}
\label{gdp} \\
\text{Inflation}& \text{:} & & & b_{\pi }(L)\pi _{t}& =b_{y}\tilde{y}%
_{t-1}+\varepsilon _{t}^{\pi }  \label{AS} \\
\text{Output gap}& \text{:} & & & a_{y}(L)\tilde{y}_{t}& =a_{r}\left(
L\right) [r_{t}-4g_{t}-z_{t}]+\varepsilon _{t}^{\tilde{y}}  \label{IS} \\
\text{Output trend}& \text{:} & & & y_{t}^{\ast }& =y_{t-1}^{\ast
}+g_{t-1}+\varepsilon _{t}^{y^{\ast }}~\hsp[15]  \label{y*} \\
\text{Trend growth}& \text{:} & & & g_{t-1}& =g_{t-2}+\varepsilon _{t-1}^{g}
\label{g} \\
\text{Other factor}& \text{:} & & & z_{t-1}& =z_{t-2}+\varepsilon _{t-1}^{z},
\label{z}
\end{alignat}%
\esq where the terms $b_{\pi }(L)=(1-b_{\pi }L-(1-b_{\pi
})(L^{2}+L^{3}+L^{4}))$, $a_{y}(L)=(1-a_{y,1}L-a_{y,2}L^{2})$, and $a_{r}(L)=%
\tfrac{a_{r}}{2}(L+L^{2})$ are lag polynomials that capture the dynamics in
inflation $\pi _{t}$, the output gap $\tilde{y}_{t}$, and the real rate
cycle $\tilde{r}_{t}=(r_{t}-4g_{t}-z_{t})$, and $L$ denotes the lag
operator. Output $y_{t}$ is constructed as 100 times the log of real GDP, $%
\pi _{t}$ is annualized quarter-on-quarter PCE inflation, and the real
interest rate $r_{t}$ is computed as $r_{t}=i_{t}-\pi _{t}^{e},$ where $%
i_{t} $ is the federal funds rate, and expected inflation is approximated by
$\pi _{t}^{e}=\frac{1}{4}\sum_{i=0}^{3}\pi _{t-i}$. The natural rate of
interest $r_{t}^{\ast }$ is defined as the sum of annualized trend growth $%
4g_{t}$ and other factor $z_{t}$, each of which follow first order
integrated processes. The error terms $\varepsilon _{t}^{\ell }$ $\left(
\forall \ell =\{\pi ,\tilde{y},y^{\ast }\hsp[-1],g,z\}\right) $ in \ref%
{eq:hlw} are assumed to be $i.i.d.$, mutually uncorrelated, and with
time-invariant standard deviations denoted by $\sigma _{\ell }$.

HLW argue that due to pile-up at zero problems with Maximum Likelihood
Estimation (MLE) of $\sigma _{g}$ and $\sigma _{z}$ in \ref{eq:hlw}, these
estimates are \textit{`likely to be biased towards zero'} (\cite%
{holston.etal:2017}, p. S64). To avoid pile-up problems, HLW estimate $%
\sigma _{g}$ and $\sigma _{z}$ indirectly from the signal-to-noise ratios $%
\lambda _{g}=\frac{\sigma _{g}}{\sigma _{y^{\ast }}}$ and $\lambda _{z}=%
\frac{a_{r}\sigma _{z}}{\sigma _{\tilde{y}}}$ computed in two preliminary
stages (referred to as Stage 1 and Stage 2) using SW's MUE. In the sections
below, I\ briefly outline how MUE is implemented in SW, before describing
the procedure that HLW adopt in Stage 2. This description is necessary to
understand that these differences in implementation lead to very different
estimates of $\lambda _{z}$, other factor $z_{t}$, and ultimately the
natural rate $r_{t}^{\ast }$. Because I do not encounter any pile-up at zero
problems with MLE of $\sigma _{g}$, only HLW's Stage 2 model and
implementation of MUE is described in what follows.\footnote{%
A referee has pointed out that MLE of $\sigma _{g}$ in the full model in
\ref{eq:hlw} may not generate pile-up at zero problems in the sample that I
consider, but may well have in the sample of data used when the estimation
framework was proposed in \cite{laubach.williams:2003}, ie., with data up to
2002:Q2. In \autoref{fig:US_rec_sigma_g} I plot recursive ML estimates of $%
\sigma _{g}$ and $\sigma _{z}$ from HLW's full model in \ref{eq:hlw} with
the sample ending in 1987:Q2 up to 2019:Q4, with one quarter update
increments. The data were downloaded from FRED\ on the $28^{th}$ of May
2020. The MLE$\ $of $\sigma _{g}$ never shrinks to zero. The MLE of $\sigma
_{z}$ does shrink to zero in every sample up to mid 2018. Although\ I do not
have the vintage of data from their 2003 paper, estimating $\sigma _{g}$ by
MLE does not generate any pile-up at zero problems in currently available
data.}

\subsection{Median Unbiased Estimation in SW and HLW's Stage 2 model}

\cite{stock.watson:1998} introduced MUE in the context of a local-level
model for the estimation of US\ per capita trend growth, taking the form:\bsq%
\label{eq:sw98}
\begin{align}
GY_{t}& =\beta _{t}+u_{t}  \label{eq:sw1} \\
\Delta \beta _{t}& =(\lambda /T)\eta _{t}  \label{eq:swRW} \\
a(L)u_{t}& =\varepsilon _{t},  \label{sw:AR4}
\end{align}%
\esq where $\eta _{t}$ and $\varepsilon _{t}$ are two $i.i.d.$ and mutually
uncorrelated disturbance terms. They derived the MUE parameter of interest $%
\lambda $ to be: \textquotedblleft \textit{... }$T$\textit{\ times \textit{%
the }ratio of the long-run standard deviation of }$\Delta \beta _{t}$\textit{%
\ to the long run standard deviation of }$u_{t}$\textit{.}%
\textquotedblright\ (\cite{stock.watson:1998}, p. 351, right column, top of
the page, with $T$ denoting the sample size). Re-arranging, gives the
following definition of the signal-to-noise ratio parameter of interest:%
\begin{equation}
\lambda _{z}=\frac{\lambda }{T}=\frac{\bar{\sigma}(\Delta \beta _{t})}{\bar{%
\sigma}(u_{t})}=\frac{\sigma _{\Delta \beta }}{\sigma _{\varepsilon }/a(1)},
\label{eq:lambda}
\end{equation}%
where $a(L)$ is an AR(4)$\ $lag polynomial and $\bar{\sigma}(\cdot )$
denotes the long-run standard deviation. Equation \ref{eq:lambda} relates $%
\lambda _{z}=\lambda /T$ to the signal-to-noise ratio $\frac{\bar{\sigma}%
(\Delta \beta _{t})}{\bar{\sigma}(u_{t})}$ in the local-level model in \ref%
{eq:sw98}.

In \cite{holston.etal:2017}, the signal-to-noise ratio $\lambda _{z}=\frac{%
a_{r}\sigma _{z}}{\sigma _{\tilde{y}}}$ is used in the estimation of the
full model in \ref{eq:hlw}. To see algebraically why $\lambda _{z}=\frac{%
a_{r}\sigma _{z}}{\sigma _{\tilde{y}}}$ in HLW's model, assume for now that
the latent cycle and trend growth variables $\tilde{y}_{t}$ and $g_{t}$, as
well as all the parameters $a_{y,1}$, $a_{y,2}$ and $a_{r}$ in \ref{eq:hlw}
are known. The objective is to obtain an estimate of the standard deviation
of $\varepsilon _{t}^{z}$ ($\sigma _{z}$) using SW's MUE. To do so, one
needs to formulate a local-level model involving the output gap in \ref{IS}
and the equation for the latent process $z_{t}$ in \ref{z}. This formulation
is analogous to the trend growth specification in \ref{eq:sw98} and takes
the following form:\bsq\label{MUE2}%
\begin{align}
\overbrace{a_{y}(L)\tilde{y}_{t}-a_{r}(L)[r_{t}-4g_{t}]}^{\text{analogue to }%
GY_{t}\text{ in \ref{eq:sw1}}}& =\overbrace{-a_{r}(L)z_{t}}^{%
\mathclap{\substack{\text{analogue to } \\[1pt] \beta _{t} \text{ in
\ref{eq:sw1}}}}}~+\overbrace{\varepsilon _{t}^{\tilde{y}}}^{%
\mathclap{\substack{\text{analogue to} \\[1pt] u _{t} \text{ in
\ref{eq:sw1}}}}}  \label{MUE2a} \\
\underbrace{-a_{r}(L)\Delta z_{t}}_{\mathclap{\substack{\text{analogue to }
\\[1pt] \Delta\beta _{t} \text{ in \ref{eq:swRW}}}}}& =\underbrace{%
-a_{r}(L)\varepsilon _{t}^{z}}_{\mathclap{\substack{\text{analogue to }
\\[1pt] (\lambda /T)\eta _{t} \text{ in \ref{eq:swRW}}}}},  \label{MUE2b}
\end{align}%
\esq where $-a_{r}(L)z_{t}$ is the counterpart to the time-varying mean
process $\beta _{t}$ in \ref{eq:sw98}. The signal-to-noise ratio
corresponding to \ref{eq:lambda} for the local-level model specification of
the relevant equation of HLW's model in \ref{MUE2} is then:%
\begin{equation}
\lambda _{z}=\frac{\lambda }{T}=\frac{\bar{\sigma}(\Delta \beta _{t})}{\bar{%
\sigma}(\varepsilon _{t}^{\tilde{y}})}=\frac{\bar{\sigma}(-a_{r}(L)\Delta
z_{t})}{\sigma _{\tilde{y}}}=\frac{a_{r}(1)\sigma _{z}}{\sigma _{\tilde{y}}}=%
\frac{a_{r}\sigma _{z}}{\sigma _{\tilde{y}}},  \label{mue2_ratio}
\end{equation}%
due to $a_{r}(1)=\frac{a_{r}}{2}(1+1^{2})=a_{r}$, and $\bar{\sigma}%
(\varepsilon _{t}^{\tilde{y}})=\sigma _{\tilde{y}}$, since $\varepsilon
_{t}^{\tilde{y}}$ is serially uncorrelated by assumption. The last term in %
\ref{mue2_ratio} gives HLW's Stage 2 signal-to-noise ratio $\lambda _{z}=%
\frac{a_{r}\sigma _{z}}{\sigma _{\tilde{y}}}$. Once $\lambda _{z}$ has been
estimated by MUE, $\sigma _{z}$ is replaced by $\frac{\hat{\lambda}%
_{z}\sigma _{\tilde{y}}}{a_{r}}$ in the full model's likelihood function,
and $\hat{\lambda}_{z}$ is held fixed at the MUE\ point estimate in the
estimation of the remaining parameters of the model in equation \ref{eq:hlw}.

\subsection{The empirical Stage 2 model in HLW and the Correct specification}

In the derivation of the algebraic relationship between $\lambda _{z}$ and $%
\frac{a_{r}\sigma _{z}}{\sigma _{\tilde{y}}}$ in \ref{mue2_ratio} it was
assumed that the latent cycle and trend growth variables $\tilde{y}_{t}$ and
$g_{t}$, as well as all the parameters $a_{y,1}$, $a_{y,2}$ and $a_{r}$ in %
\ref{MUE2} are known. In practice, however, these are unknown and will need
to be replaced by estimates obtained from a preliminary model, called the `%
\textit{Stage 2}' model in HLW. This Stage 2 model is a restricted form of
the full model in \ref{eq:hlw} and is needed to compute estimates of the
unknown parameters and latent states of the empirical counterpart to the
local-level model in \ref{MUE2}. As such, the Stage 2 model should simply be
defined as the full model in \ref{eq:hlw}, but with $z_{t}$ excluded from
the output gap relation in \ref{IS}, and the equation for $z_{t}$ in \ref{z}
entirely removed from the model. This `\textit{Correct Stage 2}' model,
shown in the left column block in \ref{eq:S2} below, is consistent with the
full model's output gap definition in \ref{IS} and yields HLW's
signal-to-noise ratio $\lambda _{z}=\frac{a_{r}\sigma _{z}}{\sigma _{\tilde{y%
}}}$, as demonstrated algebraically in \ref{MUE2} and \ref{mue2_ratio} above.

Instead of formulating the Stage 2 model in this way, HLW modify the output
gap equation to contain only one lag of trend growth in the model and
further add an intercept term, with the parameter on the lagged trend growth
term not restricted to be $-4a_{r}$.\footnote{%
The inclusion of the intercept term has been defended by HLW, arguing: "%
\textit{A constant is needed in this equation because excluding it would
impose a sample mean of zero for the variable }$z_{t}$\textit{, which
captures the movements in the natural rate of interest not related to trend
GDP growth. Such a restriction is neither implied by the model (z is a
random walk and therefore the sample mean need not be zero), nor is it
supported by the data.}" I show in the Stage 2 estimates reported in %
\tabsref{tab:US_Stage2}, \nref{tab:EA_Stage2}, \nref{tab:UK_Stage2} and %
\nref{tab:CA_Stage2} that the intercept term is in fact \emph{not} supported
by the data\ (see the last column under the heading `Correct$~+a_{0}$'). The
largest increase in the log-likelihood function is $0.3847$ for the US. With
one degree of freedom, $2\times 0.3847$, yields an upper Chi-square $p-$%
value of $0.3804$. A constant is not needed in the correct Stage 2 model
shown in the left column block of \ref{eq:S2}. For the Euro Area, the UK and
Canada, the increases in the log-likelihoods are even smaller at $0.0414$, $%
0.0146$, and $0.0002$.} These two different Stage 2 model specifications are
contrasted in the left and right column blocks of \ref{eq:S2} below.%
\footnote{%
The trend growth equation in \ref{trnd} is also misspecified due to $g_{t-2}$
instead of $g_{t-1}$ being included in the relation, making $\mathring{%
\varepsilon _{t}}^{y^{\ast }}\hsp[-1]=\varepsilon _{t}^{y^{\ast
}}+g_{t-1}-g_{t-2}=\varepsilon _{t}^{y^{\ast }}+\varepsilon _{t-1}^{g}$,
that is, an MA(1)\ process. Due to the additional $\varepsilon _{t-1}^{g}$
term in the $y_{t}^{\ast }$ (trend) equation, the covariance matrix of the
error terms of the state vector will not be diagonal anymore. Since the
focus here is on the output gap misspecification, I do not discuss this any
further.} \bsq\label{eq:S2}%
\begin{align}
& \hsp[-10]\text{\textbf{Correct Stage 2 model}} & & \hsp[-05]\text{%
\bfrr{HLW's (misspecified) Stage 2 model}}  \notag \\[0.02in]
y_{t}& =y_{t}^{\ast }+\tilde{y}_{t} & y_{t}& =y_{t}^{\ast }+\tilde{y}_{t} \\
b_{\pi }(L)\pi _{t}& =b_{y}\tilde{y}_{t-1}+\varepsilon _{t}^{\pi } & \hsp[3]%
b_{\pi }(L)\pi _{t}& =b_{y}\tilde{y}_{t-1}+\varepsilon _{t}^{\pi } \\
a_{y}(L)\tilde{y}_{t}& =a_{r}\left( L\right) [r_{t}-4g_{t}]+\mathring{%
\varepsilon}_{t}^{\tilde{y}} & \hsp[5]a_{y}(L)\tilde{y}_{t}& =\rred{a_{0}}%
+a_{r}\left( L\right) r_{t}+\rred{a_{g}g_{t-1}}+\mathring{\varepsilon}_{t}^{%
\tilde{y}}  \label{cycl} \\
y_{t}^{\ast }& =y_{t-1}^{\ast }+g_{t-1}+\varepsilon _{t}^{y^{\ast }} &
y_{t}^{\ast }& =y_{t-1}^{\ast }+g_{t-2}+\mathring{\varepsilon _{t}}^{y^{\ast
}}  \label{trnd} \\
g_{t-1}& =g_{t-2}+\varepsilon _{t-1}^{g} & g_{t-1}& =g_{t-2}+\varepsilon
_{t-1}^{g} \\[-7mm]
& & &  \notag
\end{align}%
\esq The error terms $\mathring{\varepsilon}_{t}^{\tilde{y}}$ corresponding
to the two output-gaps in \ref{cycl} are, respectively:\vsp[-2]\bsq\label%
{eq:S2_errors}
\begin{align}
& \hsp[-7]\text{\textbf{Correct Stage 2 model}} & & \hsp[17]\text{%
\bfrr{HLW's (misspecified) Stage 2 model}}  \notag \\
\mathring{\varepsilon}_{t}^{\tilde{y}}& =-a_{r}(L)z_{t}+\varepsilon _{t}^{%
\tilde{y}} & \hsp[5]\mathring{\varepsilon}_{t}^{\tilde{y}}& =\underbrace{%
-a_{r}(L)4g_{t}-a_{r}(L)z_{t}+\varepsilon _{t}^{\tilde{y}}}_{\text{missing
true model part}}-\underbrace{(a_{0}+a_{g}g_{t-1})}_{\text{added Stage 2 part%
}} \\
& & & =\underbrace{-a_{r}(L)z_{t}+\varepsilon _{t}^{\tilde{y}}}_{%
\mathclap{\substack{\text{required terms from} \\[0pt]  \text{true model}}}%
}-\underbrace{\left[ a_{0}+a_{g}g_{t-1}+a_{r}(L)4g_{t}\right] }_{\text{%
\rred{unnecessary terms}}} \\
& & & =-a_{r}(L)z_{t}+\underbrace{\varepsilon _{t}^{\tilde{y}}-[a_{0}+\tfrac{%
(a_{g}+4a_{r})}{2}(g_{t-1}+g_{t-2})+\tfrac{a_{g}}{2}\varepsilon _{t-1}^{g}]}%
_{\rred{\mathring{\nu}_{t}^{\tilde{y}}}}  \label{nu} \\[-2mm]
& & & =-a_{r}(L)z_{t}+\rred{\mathring{\nu}_{t}^{\tilde{y}}}.
\end{align}%
\esq\vsp[-10]

To see what theoretical signal-to-noise ratio $\lambda _{z}$ can be
recovered from HLW's (misspecified) Stage 2 model specification in the right
column of \ref{eq:S2_errors}, one needs to go through the same algebraic
steps as in equations \ref{MUE2} and \ref{mue2_ratio} above. That is, assume
that $\tilde{y}_{t}$ and $g_{t}$, as well as $a_{y,1}$, $a_{y,2}$, $a_{r}$, $%
a_{0}$, $a_{g}$ are known (or have been estimated). Then, formulating once
again a local-level model analogous to \ref{MUE2}, but now for HLW's
(misspecified) Stage 2 model, yields:\bsq\label{S2wrong}%
\begin{align}
\overbrace{a_{y}(L)\tilde{y}_{t}-a_{0}-a_{r}(L)r_{t}-a_{g}g_{t-1}}^{\text{%
misspecified analogue to }GY_{t}}& =\overbrace{-a_{r}(L)z_{t}}^{%
\mathclap{~\text{analogue to } \beta _{t} }}~+\overbrace{\mathring{\nu}_{t}^{%
\tilde{y}}}^{\mathclap{\substack{\text{analogue} \\[1pt] \text{to } u _{t} }}%
}  \label{S2wrong_a} \\
\underbrace{-a_{r}(L)\Delta z_{t}}_{\mathclap{\substack{\text{analogue}
\\[1pt] \text{to } \Delta\beta _{t}}}}& =\underbrace{-a_{r}(L)\varepsilon
_{t}^{z}}_{\mathclap{\substack{\text{analogue} \\[1pt] \text{to } (\lambda
/T)\eta _{t} }}},  \label{S2wrong_b}
\end{align}%
\esq where
\begin{equation}
\mathring{\nu}_{t}^{\tilde{y}}=\varepsilon _{t}^{\tilde{y}}-[a_{0}+\tfrac{%
(a_{g}+4a_{r})}{2}(g_{t-1}+g_{t-2})+\tfrac{a_{g}}{2}\varepsilon _{t-1}^{g}]
\label{nu_0}
\end{equation}%
in \ref{S2wrong_a} is the (misspecified) error term corresponding to $%
\varepsilon _{t}^{\tilde{y}}$ in the local-level model in \ref{MUE2}.%
\footnote{%
Observe that $g_{t-1}=\frac{1}{2}(g_{t-1}+g_{t-1})=\frac{1}{2}%
(g_{t-1}+g_{t-2}+\varepsilon _{t-1}^{g})$ so that $a_{g}g_{t-1}=\frac{a_{g}}{%
2}(g_{t-1}+g_{t-2})+\frac{a_{g}}{2}\varepsilon _{t-1}^{g}.$} The resulting
signal-to-noise ratio is then:%
\begin{equation}
\lambda _{z}=\frac{\lambda }{T}=\frac{\bar{\sigma}(-a_{r}(L)\Delta z_{t})}{%
\bar{\sigma}(\rred{\mathring{\nu}_{t}^{\tilde{y}}})}=\frac{a_{r}(1)\sigma
_{z}}{\bar{\sigma}(\rred{\mathring{\nu}_{t}^{\tilde{y}}})}=\frac{a_{r}\sigma
_{z}}{\bar{\sigma}(\rred{\mathring{\nu}_{t}^{\tilde{y}}})},
\label{Lambda_z_correct}
\end{equation}%
which now requires the evaluation of the long-run standard deviation of $%
\rred{\mathring{\nu}_{t}^{\tilde{y}}}$ in the denominator of \ref%
{Lambda_z_correct}. Even if $(a_{g}+4a_{r})=0$ in \ref{nu_0}, the long-run
standard deviation of $\rred{\mathring{\nu}_{t}^{\tilde{y}}}$ will depend on
$\tfrac{a_{g}}{2}\sigma _{g}$ due to the extra $\tfrac{a_{g}}{2}\varepsilon
_{t-1}^{g}$ term in $\rred{\mathring{\nu}_{t}^{\tilde{y}}}$, giving:
\begin{equation}
\lambda _{z}=\frac{\lambda }{T}=\frac{a_{r}\sigma _{z}}{(\sigma _{\tilde{y}%
}+a_{g}\sigma _{g}/2)}.  \label{Lz00}
\end{equation}%
As can be seen from \ref{Lz00}, MUE applied to HLW's (misspecified) Stage 2
model cannot recover the ratio of interest $\lambda _{z}=\frac{a_{r}\sigma
_{z}}{\sigma _{\tilde{y}}}$. Estimating the full model in \ref{eq:hlw} with $%
\sigma _{z}$ replaced by $\frac{\lambda _{z}\sigma _{\tilde{y}}}{a_{r}}$ in
the Kalman Filter recursion that builds up the likelihood function is thus
inconsistent with the signal-to-noise ratio obtained from HLW's
(misspecified) Stage 2 model.

\subsection{Structural break regressions in MUE\label{sec:break}}

The previous section showed algebraically that HLW's (misspecified) Stage 2
model cannot recover the ratio of interest $\lambda _{z}=\frac{a_{r}\sigma
_{z}}{\sigma _{\tilde{y}}}$ from MUE. I now contrast SW's and HLW's
implementations of the structural break regressions used as the auxiliary
model in MUE.

Note that MUE is fundamentally an indirect estimator. A structural break
test statistic is used to recover the parameter of interest, which is the
signal-to-noise ratio $\lambda _{z}$. SW do not only provide the theory
behind MUE, but also supply look-up values (Table 3 on page 354 in \cite%
{stock.watson:1998}) that convert a set of structural break test statistics
into Median Unbiased estimates of $\lambda $ (dividing by the sample size $T$
gives $\lambda _{z}$). These look-up values were constructed by simulation,
and are valid for the local-level model and the structural break tests as
implemented in SW.\footnote{%
As a reminder, the look-up table for MUE\ of $\lambda $ on page 354 in \cite%
{stock.watson:1998} was constructed by simulating $T=500$ observations from
the local-level model for increasing values of $\lambda $ from $0,1,\ldots
,30$, and then testing the observed series for a structural break in the
unconditional mean as in \ref{eq:sw_break}, that is, without any other
`extra regressors'. This process was repeated 5000 times. The resulting 5000
values from the structural break test statistics were then ordered, with the
median of those being the MUE. When employing MUE empirically, one works the
other way around and infers the value of $\lambda $ from the structural
break test statistics.}

\cite{stock.watson:1998} construct the structural break tests as follows.
First, the GDP growth variable $GY_{t}$ in \ref{eq:sw1} is filtered by
fitting an AR(4) model to capture the dynamics of $u_{t}$.\footnote{%
See the implementation in \SW GAUSS\ file \texttt{TST\_GDP1.GSS} which is
available from Mark Watson's homepage at %
\url{http://www.princeton.edu/~mwatson/ddisk/tvpci.zip}, in particular, lines
39 to 66 which AR(4)\ filter the GDP\ growth data, and lines 68 to 83 which
then implement the \cite{chow:1960} type structural break tests to the AR(4)
filtered data.} Then, the AR(4)\ filtered $GY_{t}$ series (which I denote by
$\hat{a}(L)GY_{t}$ below)\ is tested for a structural break in the
unconditional mean by running the following dummy variable regression for
each potential break point $\tau \in \lbrack \tau _{0},\tau _{1}]$:
\newline
\vsp[-3]
\begin{equation}
\hat{a}(L)GY_{t}=\zeta _{0}+\zeta _{1}D_{t}(\tau )+\epsilon _{t},
\label{eq:sw_break}
\end{equation}%
where $\hat{a}(L)$ is the estimated counterpart to the AR(4)$\ $lag
polynomial $a(L)$ in \ref{sw:AR4}, $D_{t}(\tau )$ is a dummy variable that
is equal to $1$ if $t>\tau ,$ and $0$ otherwise, and $\tau =\{\tau _{0},\tau
_{0}+1,\tau _{0}+2,\ldots ,\tau _{1}\}$ is an index (or sequence) of grid
points between endpoints $\tau _{0}$ and $\tau _{1}$.\footnote{\cite%
{stock.watson:1998} set these endpoints at the $15^{th}$ and $85^{th}$
percentiles of the sample size $T$, that is, $\tau _{0}=0.15T$ and $\tau
_{1}=0.85T$. More precisely, $\tau _{0}$ is computed as $\mathtt{%
floor(0.15\ast T)}$ and $\tau _{1}$ as $T-\tau _{0}$ in their GAUSS code. In
HLW, these are set at $\tau _{0}=4$ and $\tau _{1}=T-4$, respectively.} The
(sequence of) $F$ statistics $\left( \{F(\tau )\}_{\tau =\tau _{0}}^{\tau
_{1}}\right) $ on the $\hat{\zeta}_{1}(\tau )$ point estimates is then
utilized in the computation of the following structural break test
statistics:\vsp[-1]\bsq\label{eq:breakTests}%
\begin{align}
\mathrm{MW}& =\frac{1}{N_{\tau }}\sum\limits_{\tau =\tau _{0}}^{\tau
_{1}}F(\tau )  \label{MW} \\[2mm]
\mathrm{EW}& =\ln \bigg(\frac{1}{N_{\tau }}\sum_{\tau =\tau _{0}}^{\tau
_{1}}\exp \left\{ \frac{1}{2}F(\tau )\right\} \bigg)  \label{EW} \\[4mm]
\mathrm{QLR}& =\max_{\tau \in \lbrack \tau _{0},\tau _{1}]}\{F(\tau
)\}_{\tau =\tau _{0}}^{\tau _{1}},  \label{QLR}
\end{align}%
\esq where MW and EW are, respectively, \cites{andrews.ploberger:1994} mean
Wald and exponential Wald tests, and QLR is \cites{quandt:1960} Likelihood
ratio test.\ SW also compute \NL$L$ statistic, which is constructed directly
from the sum of squared cumulative sums of the demeaned $\hat{a}(L)GY_{t}$
series, and therefore does not require a structural break regression
involving dummy variables as in \ref{eq:sw_break}.\footnote{%
In our setting, \NL$L$ statistic provides a consistency check on the break
test implementation, as it is not affected by how the structural break tests
are implemented.} Look-up Table 3 in \cite{stock.watson:1998} gives the
mapping between the various structural break test statistics and $\lambda $
values.

Observe how the structural break regressions in \ref{eq:sw_break} are
implemented. The regressand, the left-hand side variable $\hat{a}(L)GY_{t}$
in \ref{eq:sw_break}, is constructed only once outside the dummy variable
regression loop and there are no other `extra regressors' on the right-hand
side, ie., only the break dummy is included on the right-hand side. To make
this last point clear, the break regressions are \textit{not} estimated as:%
\begin{equation}
GY_{t}=\underbrace{a_{1}GY_{t-1}+a_{2}GY_{t-2}+a_{3}GY_{t-3}+a_{4}GY_{t-4}}_{%
\text{extra regressors}}+\zeta _{0}+\zeta _{1}D_{t}(\tau )+\epsilon _{t},
\label{eq:sw_break_false}
\end{equation}%
ie., with $GY_{t}$ as the left-hand side variable, and extra regressors $%
\left\{ GY_{t-i}\right\} _{i=1}^{4}$ added to the right-hand side of the
regression in \ref{eq:sw_break}. These two different forms of the break test
implementations in \ref{eq:sw_break} and \ref{eq:sw_break_false} lead to
vastly different sequences of $F(\tau )$ statistics on $\hat{\zeta}_{1}(\tau
)$. In \autoref{fig:SW_Fstats} I provide a visual illustration of how
different the resulting $F(\tau )$ sequences computed from the two versions
of the structural break regressions are in the context of SW's study on
trend growth. The implementation in \ref{eq:sw_break_false} results in a
sequence that is substantially larger at values between 7 to 8, while SW's
original implementation in \ref{eq:sw_break} yields estimates in the 0 to
approximately 3 range. The MW, EW and QLR structural break statistics
defined in \ref{eq:breakTests} corresponding to the different $F(\tau )$
sequences can differ by a factor of (almost) up to 10.\footnote{%
Specifically, these are:\ $\{0.1103,0.0987,0.0250\}$ and $%
\{0.4461,0.3426,0.2342\}$ for the MW, EW and QLR tests, from the break
regressions in \ref{eq:sw_break} and \ref{eq:sw_break_false}, respectively.}
Since the break statistics are used together with the look-up values in
Table 3 of SW to arrive at the MUE\ of $\lambda $, any differences in these
will directly impact the estimates of $\lambda _{z}=\lambda /T$.

Now HLW's implementation of the structural break regressions is of the
second form, that is, as in \ref{eq:sw_break_false}. What makes matters
worse is that these structural break regressions are implemented on HLW's
(misspecified) Stage 2 model shown in the right column of \ref{eq:S2}, which
spuriously amplifies the sequence of $F(\tau)$ statistics even further.
Specifically, HLW obtain $\{F(\tau )\}_{\tau =\tau _{0}}^{\tau _{1}}$ on $%
\hat{\zeta}_{1}(\tau ) ~ \forall \tau \in \lbrack \tau _{0},\tau _{1}]$ from
the following break regression:%
\begin{equation}
\hat{\tilde{y}}_{t|T}=a_{0}+\underbrace{a_{1}\hat{\tilde{y}}_{t-1|T}+a_{2}%
\hat{\tilde{y}}_{t-2|T}+a_{r}(r_{t-1}+r_{t-2})/2+a_{g}\hat{g}_{t-1|T}}_{%
\text{extra regressor}}+\zeta _{1}D_{t}(\tau )+\epsilon _{t}  \label{d_hlw}
\end{equation}%
rather than from:%
\begin{equation}
\underbrace{\hat{a}_{y}(L)\hat{\tilde{y}}_{t|T}-\hat{a}_{0}-\hat{a}%
_{r}(L)r_{t}-\hat{a}_{g}\hat{g}_{t-1|T}}_{\text{observed analogue to }GY_{t}%
\text{ in \ref{S2wrong_a}}}=\zeta _{0}+\zeta _{1}D_{t}(\tau )+\epsilon _{t}.
\label{d_true}
\end{equation}%
For the correct Stage 2 model, these would be:%
\begin{equation}
\hat{\tilde{y}}_{t|T}=\underbrace{a_{1}\hat{\tilde{y}}_{t-1|T}+a_{2}\hat{%
\tilde{y}}_{t-2|T}+a_{r}(r_{t-1}+r_{t-2}-4[\hat{g}_{t-1|T}+\hat{g}%
_{t-2|T}])/2}_{\text{extra regressor}}+\zeta _{0}+\zeta _{1}D_{t}(\tau
)+\epsilon _{t},  \label{d_hlw0}
\end{equation}%
and:
\begin{equation}
\underbrace{\hat{a}_{y}(L)\hat{\tilde{y}}_{t|T}-\hat{a}_{r}(L)[r_{t}-4\hat{g}%
_{t-1|T}]}_{\text{observed analogue to }GY_{t}\text{ in \ref{MUE2a}}}=\zeta
_{0}+\zeta _{1}D_{t}(\tau )+\epsilon _{t},  \label{d_true0}
\end{equation}%
where $\hat{a}_{y,1}$, $\hat{a}_{y,2}$, $\hat{a}_{r}$, $\hat{a}_{0}$, $\hat{a%
}_{g}$ are the corresponding estimated Stage 2 coefficients, and $\{\hat{%
\tilde{y}}_{t-i|T}\}_{i=0}^{2}$ and $\{\hat{g}_{t-i|T}\}_{i=1}^{2}$ are the
Kalman smoothed estimates of the output gap $\tilde{y}_{t}$ and trend growth
$g_{t}$, respectively.

Since all Stage 2 model parameters as well as latent states have already
been estimated from the full sample of data before implementing the
structural break regressions needed for MUE, there is no reason \textit{not }%
to formulate the left-hand side variable in the local-level model to be
consistent with the specification in \ref{MUE2a} used to (theoretically)
derive the signal-to-noise ratio to be $\lambda _{z}=\frac{a_{r}\sigma _{z}}{%
\sigma _{\tilde{y}}}$. It is the variation in the unconditional mean of the
left-hand side variable $(a_{y}(L)\tilde{y}_{t}-a_{r}(L)[r_{t}-4g_{t}])$ in %
\ref{MUE2a} that needs to be tested for a structural break, and not the
variation in the mean of $\tilde{y}_{t}$, conditional on $\left\{ \tilde{y}%
_{t-i},r_{t-i},g_{t-i}\right\} _{i=1}^{2}$. For HLW's (misspecified) Stage 2
model, these two different break test implementations yield vastly different
sequences of $F(\tau )$ statistics and $\lambda _{z}$ estimates. For the
correct Stage 2 model, the differences are considerably smaller. It is
precisely the combination of HLW's (misspecified) Stage 2 model formulation
together with their modified break test implementation that affects the
parameter of interest $\lambda _{z}$ and leads to the spurious downward
trend in other factor $z_{t}$. The Stage 2 model's parameter estimates as
such are not materially affected.

\section{Empirical results \label{sec:results}}

This section provides the full empirical estimation results of the correct
Stage 2 model specification using the structural break test implementation
as in SW. HLW's estimates are also reported for comparison. Some results are
provided as supplementary information or for reasons of completeness, and do
not merit any discussion. I use the same data as described in \cite%
{holston.etal:2017}\ (see the replication files for details) with the sample
ending in 2019:Q4. Readers not interested in the Stage 2 MUE results as such
can skip directly to the plots of the smoothed natural rate $r_{t}^{\ast }$,
trend growth $g_{t}$, other factor $z_{t}$, and the output gap $\tilde{y}%
_{t} $ in \figsref{fig:US_KS}, \nref{fig:EA_KS}, \nref{fig:UK_KS} and %
\nref{fig:CA_KS} (filtered estimates are shown in in \figsref{fig:US_KF}, %
\nref{fig:EA_KF}, \nref{fig:UK_KF} and \nref{fig:CA_KF}).

In \tabsref{tab:US_Stage2}, \nref{tab:EA_Stage2}, \nref{tab:UK_Stage2} and %
\nref{tab:CA_Stage2}, the Stage 2 model parameter estimates for the US, the
Euro Area, the UK\ and Canada are reported. The first column (`HLW.R-File')
provides HLW's estimates obtained from their R-Files. The second column
(`HLW($\hat{\sigma}_{g}^{\text{\textrm{MLE}}}$)') reports estimates from
HLW's (misspecified) Stage 2 model, but with $\sigma _{g}$ estimated
directly by MLE together with the other parameters of the model. The third
column (`Correct') lists the estimates from the correct Stage 2 model
defined in the left column block of \ref{eq:S2}, with $\sigma _{g}$ again
estimated directly by MLE. The last column (`Correct + $a_{0}$') shows the
correct Stage 2 model's estimates when an additional intercept term is added
to the model.\footnote{%
Values in round brackets are implied from the Stage 1 signal-to-noise ratio
relation $\lambda _{g}=\sigma _{g}/\sigma _{y^{\ast }}$.} As can be seen,
the MLE of $\sigma _{g}$ in the Stage 2 model does not shrink to zero for
any of the estimates and is in fact larger than the one obtained from MUE\
for three of the four series. The Stage 1 model is thus not needed as an
auxiliary model to estimate $\sigma _{g}$ --- at least not for this data
set. Further, all intercept terms ($a_{0}$) in the correct Stage 2 model are
(highly) insignificant, indicating that the correct Stage 2 auxiliary model
does not need to be formulated with an intercept term.

\figsref{fig:US_Fstat}, \nref{fig:EA_Fstat}, \nref{fig:UK_Fstat} and %
\nref{fig:CA_Fstat} plot the sequences of $F(\tau )$ statistics from the
dummy variable regressions of HLW's (misspecified) Stage 2 model (top Panel
(a)) and the correct Stage 2 model (bottom Panel (b)) for both
implementations of the structural break regressions denoted by HLW\ (red
line)\ and SW\ (blue line) as defined in equations \ref{d_hlw} to \ref%
{d_true0}. These plots show that HLW's implementation of the structural
break regression leads to vastly larger sequences of $F(\tau )$ statistics
for the misspecified Stage 2 model shown in the top Panels (a), most notably
so for the US, the Euro Area, and Canada. For the same implementations of
the break tests applied to the correct Stage 2 model shown in the bottom
Panels (b), the differences in the $F(\tau )$ statistics are more subdued,
with both implementations suggesting rather low values. Note here that the
purpose of showing this comparison is to highlight the fact that it is the
combination of HLW's (misspecified) Stage 2 model and their implementation
of the structural break regression in \ref{d_hlw} that leads to the
excessively large $F(\tau )$ statistics.

In \tabsref{tab:US_Stage2_lambda_z}, \nref{tab:EA_Stage2_lambda_z}, %
\nref{tab:UK_Stage2_lambda_z} and \nref{tab:CA_Stage2_lambda_z}, the
structural break statistics corresponding to the sequences of $F(\tau )$
statistics and the resulting MUEs of $\lambda _{z}$ are reported. The tables
are arranged in two column blocks, showing the output from HLW's
implementation of the structural break regressions on the left and SW's
implementation on the right, respectively. The tables are further split into
a top and a bottom part. In the top part of the tables, the $\lambda _{z}$
estimates together with $90\%$ confidence intervals are reported, with the
bottom part showing the respective structural break test statistics and $p-$%
values in parenthesis. The column headings are the same as in the tables
showing the Stage 2 estimates, that is, `HLW.R-File', `HLW($\hat{\sigma}%
_{g}^{\text{\textrm{MLE}}}$)' and `Correct' (excluding the `Correct $+~a_{0}$%
' column). The reason for reporting the results of `HLW($\hat{\sigma}_{g}^{%
\text{\textrm{MLE}}}$)' under HLW's implementation of the structural break
tests provided in the left column block is to highlight how different the
MW, EW and QLR\ break statistics as well as MUEs of $\lambda _{z}$ are from %
\NL$L$ test. \NL$L$ statistic is (highly) insignificant for all four
countries, with $p-$values between $0.69$ and $0.945$, resulting in MUEs of $%
\lambda _{z}$ that are exactly zero.\footnote{%
A value of exactly zero for $\lambda _{z}$ is obtained whenever the
structural break test statistic is smaller than the entries corresponding to
$\lambda =0$ in the first row of Table 3 in \SW look-up values, which is
0.118 for \NL$L$ test.} The MW, EW and QLR\ tests, on the other hand,
suggest (sizeable) non-zero point estimates of $\lambda _{z}$, although the
corresponding 90\% confidence intervals indicate that these are not
statistically different from zero (HLW do not report confidence intervals on
the MUE's of $\lambda _{z}$). Under SW's implementation of the structural
break tests (right block), the tests indicate an exactly zero point estimate
of $\lambda _{z}$ for the US when implemented on HLW's version of the Stage
2 model (see the results under the heading `HLW' in the right column block).

The parameter estimates of the full model in \ref{eq:hlw} are reported in %
\tabsref{tab:US_Stage3}, \nref{tab:EA_Stage3}, \nref{tab:UK_Stage3} and %
\nref{tab:CA_Stage3}. The first column under the heading (`HLW.R-File')
lists again the estimates from HLW's R-Files as reference values. The second
column (`MLE($\sigma _{g}|\hat{\lambda}_{z}^{\text{\textrm{Correct}}}$)')
shows estimates when conditioning on the `Correct' Stage 2 MUE\ of $\lambda
_{z}$ (from EW's structural break test) and estimating $\sigma _{g}$
directly by MLE together with the other parameters of the full model in \ref%
{eq:hlw}. In the last column (`MLE($\sigma _{g},\sigma _{z}$)'), all
parameters of the full model (including $\sigma _{g}$ and $\sigma _{z}$) are
estimated by MLE (without using the Stage 2 model's $\lambda _{z}$
estimate). The following results stand out. First, as in the Stage 2 model,
the MLEs of $\sigma _{g}$ in the full model in \ref{eq:hlw} do not shrink to
zero and are again larger than their MUE based counterparts (the UK estimate
being the exception). For the US, they are $43\%$ larger; for the Euro Area,
even $83\%$ larger. Second, the MLE\ and MUE based estimates of $\sigma _{z}$
are similar when conditioning on the correct MUE\ of $\lambda _{z}$ ($\hat{%
\lambda}_{z}^{\text{\textrm{Correct}}}$), that is, when $\lambda _{z}$ is
computed from the correct Stage 2 model and when implementing the structural
break regressions as in SW. For instance, for the US, the (non-zero) ML
estimate of $\sigma _{z}$ is $0.0623$, while the MUE estimate implied from
the relation $\sigma _{z}=\hat{\lambda}_{z}^{\mathrm{Correct}}\sigma _{%
\tilde{y}}/a_{r}$ is $0.0656$. For the Euro Area, the UK and Canada, both,
the MLE and MUE\ based point estimates of $\sigma _{z}$ are zero.

The above results should not come as a surprise and are in line with the
findings in \cite{stock.watson:1998}. As a reminder, SW examine two
different ML estimators;\ one estimates the initial condition of the state
vector (referred to as MPLE\ in SW --- see Section 3.2 on pages 352 to 354
in \cite{stock.watson:1998} for details), the second one (MMLE)\ does not,
and uses instead a diffuse prior. SW show that these two MLEs have very
different pile-up at zero frequencies (see Table 1 on page 353 in SW). In
the extreme case considered in SW's simulations where the true value of $%
\lambda $ is actually 0, the MLE that does not estimate the initial
condition leads to an (at most) 16 percentage point larger pile-up at zero
frequency than MUE. MUE has a pile-up frequency of $0.5$ $(50\%)$ when $%
\lambda $ is 0. When the true $\lambda $ value is 6 (this corresponds to a $%
\lambda _{z}=\lambda /T=6/500=0.012$ which is close to the empirical
estimate of $0.013230$ for the US), the difference in pile-up frequencies
drops to 9 percentage points. Now HLW \textit{do not }estimate the initial
condition of the state vector and use rather tightly specified priors
instead.\footnote{%
This is discussed in more detail in Sections 3 and 4 in \cite{buncic:2021}.}
Getting a non-zero MLE point estimate for the US, and exactly zero estimates
for the Euro Area, the UK and Canada from both, MLE and MUE, is thus
entirely consistent with SW's simulation results.

Lastly, Kalman Filter and Smoother based estimates of the natural rate $%
r_{t}^{\ast }$, annualized trend growth $g_{t}$, other factor $z_{t}$, and
the output gap $\tilde{y}_{t}$ are shown in \figsref{fig:US_KF}, %
\nref{fig:EA_KF}, \nref{fig:UK_KF} and \nref{fig:CA_KF}, and in %
\figsref{fig:US_KS}, \nref{fig:EA_KS}, \nref{fig:UK_KS} and \nref{fig:CA_KS}%
, respectively. Estimates of other factor $z_{t}$ from the correct Stage 2
model implementation are substantially different to HLW's estimates,
particularly so for the US and the Euro Area, and somewhat less for the UK
and Canada. For the UK\ and Canada, the $z_{t}$ estimates are essentially a
horizontal line at 0, and are just below zero for the Euro\ Area. For the
US, the estimate of $z_{t}$ still shows a visible downward trend over the
full sample period. Nevertheless, relative to HLW's estimate, it is more
subdued, particulary since the global financial crisis. The difference in
these estimates is strongest at the end of the sample in 2019:Q4. Trend
growth estimates are essentially unchanged from HLW for the Euro Area, the
UK, Canada, and show a small drop in trend growth following the financial
crisis for the US. Due to the large impact of HLW's Stage 2 procedure on the
estimates of other factor $z_{t}$, the natural rate $r_{t}^{\ast }$ is
estimated to be over 100 basis points larger for the US at the end of
2019:Q4 from the correct Stage 2 model implementation. For the Euro Area,
this estimate is nearly 80 basis points larger, and for the UK and Canada,
circa 45 and 27 basis points. The magnitude of the natural rate is thus
sizeably underestimated by HLW's Stage 2 MUE\ procedure.

\section{Conclusion \label{sec:concl}}

This paper shows that \cites{holston.etal:2017} implementation of %
\cites{stock.watson:1998} Median Unbiased Estimation to determine the size
of the signal-to-noise ratio $\lambda _{z}$ cannot recover the ratio of
interest $\frac{a_{r}\sigma _{z}}{\sigma _{\tilde{y}}}$ from MUE. This
inability to recover the ratio of interest is due to a misspecification in
HLW's Stage 2 model formulation. The paper shows further that the structural
break regressions which are used as an auxiliary model in MUE are modified
from SW's original implementation. The misspecification in the Stage 2 model
--- together with the modification of the structural break regressions ---
leads to spuriously large estimates of the signal-to-noise ratio $\lambda
_{z}$, which affects the severity of the downward trend in other factor $%
z_{t}$, and thereby the estimates of the natural rate $r_{t}^{\ast }$.

The paper provides a correction to the specification of the Stage 2 model
which is consistent with the required signal-to-noise ratio $\lambda _{z}=%
\frac{a_{r}\sigma _{z}}{\sigma _{\tilde{y}}}$, and further implements the
structural break regressions following the format of SW's original
formulation to be compatible with the construction of the look-up table
values provided on page 354 in \cite{stock.watson:1998}. This correction is
quantitatively important. For the Euro Area, the UK, and Canada, the
(corrected) MUE point estimates of $\lambda _{z}$ are exactly $0$. The
downward trend in the estimates of other factor $z_{t}$ disappears entirely,
with the $z_{t}$ estimate resembling a horizontal line centered at (or very
close to) zero. For the US, the $\lambda _{z}$ point estimate shrinks from $%
0.040$ to $0.013$, resulting in a much more subdued downward trend in other
factor $z_{t}$.

The effects of HLW's Stage 2 MUE\ implementation on the estimates of the
natural rate $r_{t}^{\ast }$ are strongest at the end of the sample period
(here, in 2019:Q4); arguably when they are most needed for policy analysis.
For the US, the estimate of $r_{t}^{\ast }$ is over $100$ basis points
larger at approximately $1.5$ percentage points, than from HLW's
(misspecified) implementation of $0.48$. For the Euro Area, $r_{t}^{\ast }$
is approximately $80$ basis points larger at $1.03$ percentage points,
instead of $0.24$, while for the UK\ and Canada, the differences are more
subtle, being $45$ ($1.80$ percentage points instead of $1.35$) and $27$
basis points larger ($1.73$ percentage points instead of $1.46$).\ Estimates
of trend growth --- the second factor that determines the natural rate of
interest --- remain unchanged by the proposed correction to the Stage 2 MUE\
implementation.

\AdjustReferences\Urlmuskip=0mu plus 1mu\relax
\bibliographystyle{LongBibStyleFile}
\bibliography{DanielsBibTexLibrary}

\begin{thebibliography}{16}
\newcommand{\enquote}[1]{``#1''}
\expandafter\ifx\csname natexlab\endcsname\relax\def\natexlab#1{#1}\fi

\bibitem[\protect\citeauthoryear{Andrews and Ploberger}{Andrews and
  Ploberger}{1994}]{andrews.ploberger:1994}
\text{Andrews, Donald W.~K. and Werner Ploberger} (1994): \enquote{Optimal
  Tests when a Nuisance Parameter is Present only under the Alternative,}
  \emph{Econometrica}, \textbf{62}(6), 1383--1414.

\bibitem[\protect\citeauthoryear{Benes., Clinton, Garcia-Saltos, Johnson,
  Laxton, Manchev  and Matheson}{Benes. \it{et al.}}{2010}]{benes.etal:2010}
\text{Benes., J., K.~Clinton, R.~Garcia-Saltos, M.~Johnson, D.~Laxton,
  P.~Manchev and Troy Matheson} (2010): \enquote{Estimating Potential Output
  with a Multivariate Filter,} \emph{IMF Working Paper No. 10/285},
  International Monetary Fund.
\newblock Available from:
  \url{https://www.imf.org/external/pubs/ft/wp/2010/wp10285.pdf}.

\bibitem[\protect\citeauthoryear{Berger and Kempa}{Berger and
  Kempa}{2019}]{berger.kempa:2019}
\text{Berger, Tino and Bernd Kempa} (2019): \enquote{Testing for time variation
  in the natural rate of interest,} \emph{Journal of Applied Econometrics},
  \textbf{34}(5), 836--842.

\bibitem[\protect\citeauthoryear{Buncic}{Buncic}{2021}]{buncic:2021}
\text{Buncic, Daniel} (2021): \enquote{Econometric Issues with Laubach and
  Williams Estimates of the Natural Rate of Interest,} \emph{Sveriges Riksbank
  Working Paper No. 397}, Sveriges Riksbank.
\newblock Available from:
  \url{https://www.riksbank.se/globalassets/media/rapporter/working-papers/2019/no.-397-econometric-issues-with-laubach-and-williams-estimates-of-the-natural-rate-of-interest2.pdf}.

\bibitem[\protect\citeauthoryear{Buncic and Melecky}{Buncic and
  Melecky}{2008}]{buncic.melecky:2008}
\text{Buncic, Daniel and Martin Melecky} (2008): \enquote{An Estimated New
  Keynesian Policy Model for Australia,} \emph{The Economic Record},
  \textbf{84}(264), 1--16.

\bibitem[\protect\citeauthoryear{Chow}{Chow}{1960}]{chow:1960}
\text{Chow, Gregory~C.} (1960): \enquote{Tests of Equality between Sets of
  Coefficients in two Linear Regressions,} \emph{Econometrica}, \textbf{28}(3),
  591--605.

\bibitem[\protect\citeauthoryear{Clark}{Clark}{1987}]{clark:1987}
\text{Clark, Peter~K.} (1987): \enquote{The Cyclical Component of U.S. Economic
  Activity,} \emph{Quarterly Journal of Economics}, \textbf{102}(4), 797--814.

\bibitem[\protect\citeauthoryear{Fr\"{u}hwirth-Schnatter and
  Wagner}{Fr\"{u}hwirth-Schnatter and
  Wagner}{2010}]{fruehwirth-schnatter.wagner:2010}
\text{Fr\"{u}hwirth-Schnatter, Sylvia and Helga Wagner} (2010):
  \enquote{Stochastic model specification search for Gaussian and partial
  non-Gaussian state space models,} \emph{Journal of Econometrics},
  \textbf{154}(1), 85--100.

\bibitem[\protect\citeauthoryear{Gal{\'\i}}{Gal{\'\i}}{2015}]{gali:2015}
\text{Gal{\'\i}, Jordi} (2015): \emph{Monetary policy, inflation, and the
  business cycle: an introduction to the new Keynesian framework and its
  applications}, \emph{2nd Edition}, Princeton University Press.

\bibitem[\protect\citeauthoryear{Holston, Laubach  and Williams}{Holston \it{et
  al.}}{2017}]{holston.etal:2017}
\text{Holston, Kathryn, Thomas Laubach and John~C. Williams} (2017):
  \enquote{Measuring the Natural Rate of Interest: International Trends and
  Determinants,} \emph{Journal of International Economics},
  \textbf{108}(Supplement 1), S59--S75.

\bibitem[\protect\citeauthoryear{Kiley}{Kiley}{2020}]{kiley.etal:2020}
\text{Kiley, Michael~T.} (2020): \enquote{What Can the Data Tell us about the
  Equilibrium Real Interest Rate,} \emph{International Journal of Central
  Banking}, \textbf{16}(3), 181--209.

\bibitem[\protect\citeauthoryear{Laubach and Williams}{Laubach and
  Williams}{2003}]{laubach.williams:2003}
\text{Laubach, Thomas and John~C. Williams} (2003): \enquote{Measuring the
  Natural Rate of Interest,} \emph{Review of Economics and Statistics},
  \textbf{85}(4), 1063--1070.

\bibitem[\protect\citeauthoryear{Lewis and Vazquez-Grande}{Lewis and
  Vazquez-Grande}{2018}]{lewis.vazquez-grande:2018}
\text{Lewis, Kurt~F. and Francisco Vazquez-Grande} (2018): \enquote{Measuring
  the natural rate of interest: A note on transitory shocks,} \emph{Journal of
  Applied Econometrics}, \textbf{34}(3), 425--436.

\bibitem[\protect\citeauthoryear{Nyblom}{Nyblom}{1989}]{nyblom:1989}
\text{Nyblom, Jukka} (1989): \enquote{Testing for the Constancy of Parameters
  over Time,} \emph{Journal of the American Statistical Association},
  \textbf{84}(405), 223--230.

\bibitem[\protect\citeauthoryear{Quandt}{Quandt}{1960}]{quandt:1960}
\text{Quandt, Richard~E.} (1960): \enquote{Tests of the Hypothesis that a
  Linear Regression System obeys two Separate Regimes,} \emph{Journal of the
  American Statistical Association}, \textbf{55}(290), 324--330.

\bibitem[\protect\citeauthoryear{Stock and Watson}{Stock and
  Watson}{1998}]{stock.watson:1998}
\text{Stock, James~H. and Mark~W. Watson} (1998): \enquote{Median Unbiased
  Estimation of Coefficient Variance in a Time-Varying Parameter Model,}
  \emph{Journal of the American Statistical Association}, \textbf{93}(441),
  349--358.

\end{thebibliography}
\AddFigsTabs\AddAppendix

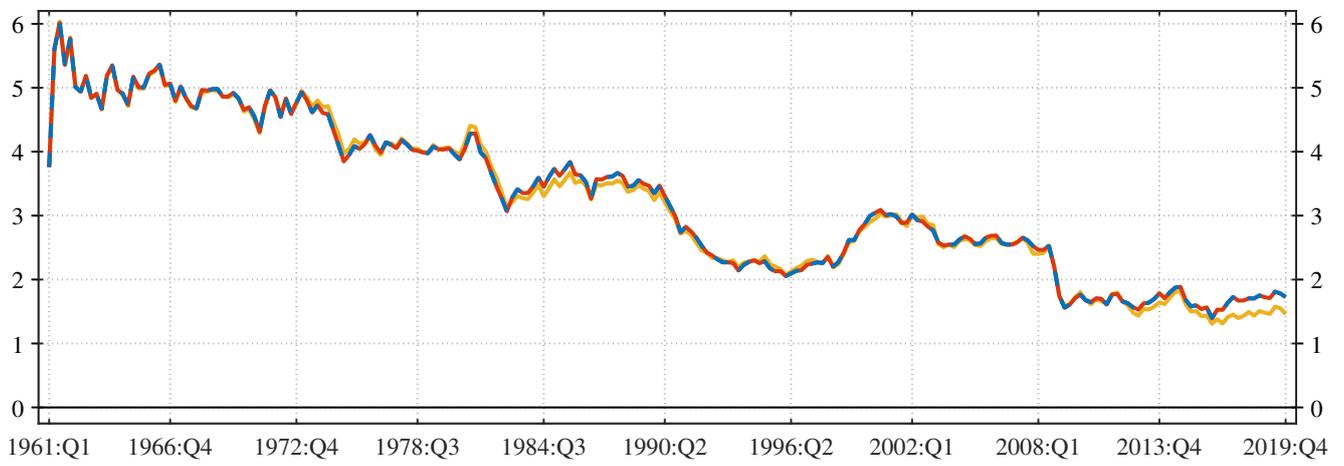
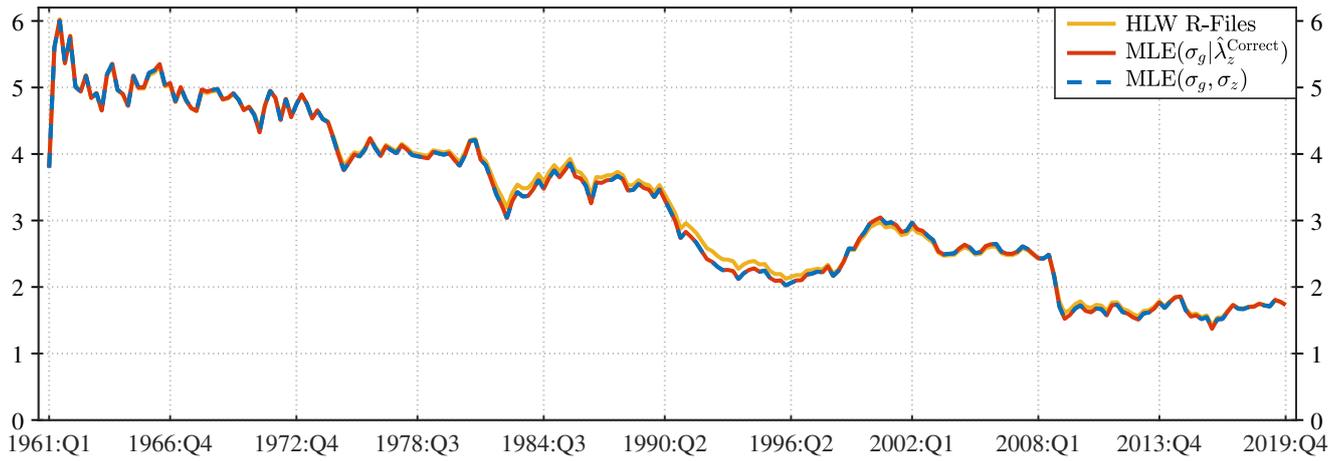
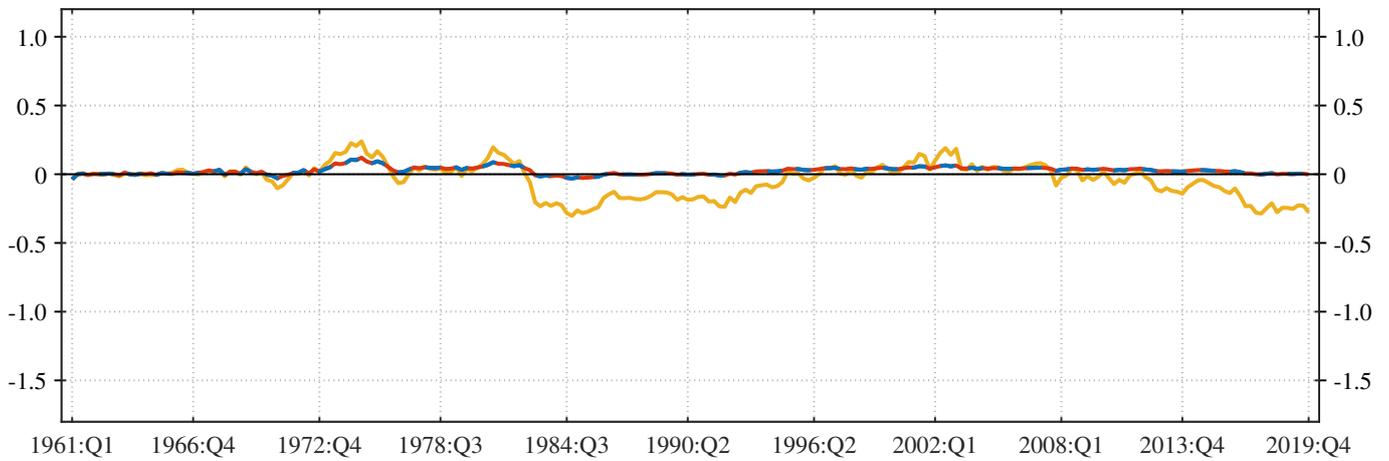
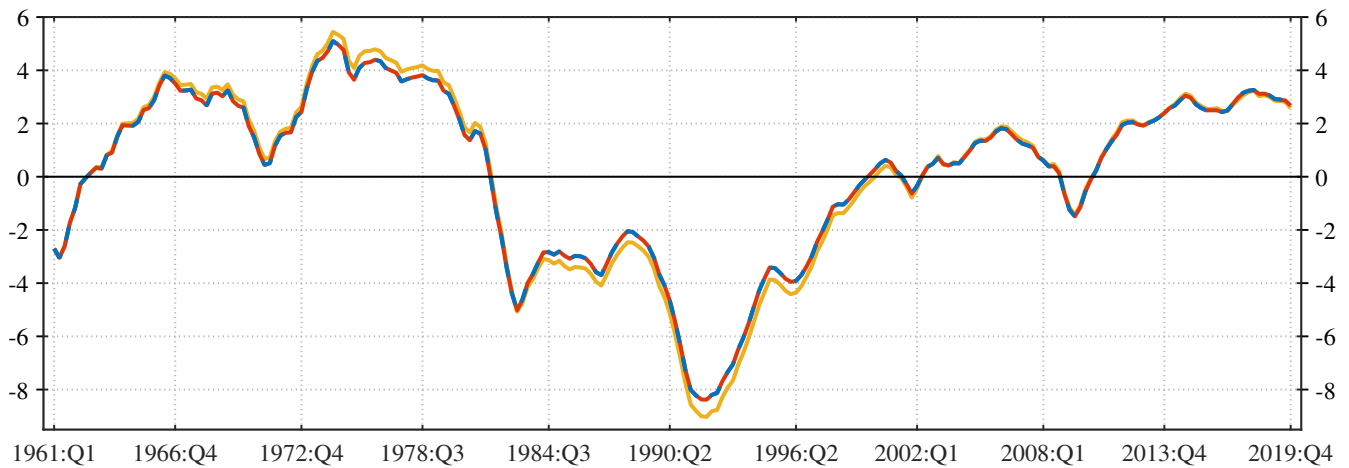
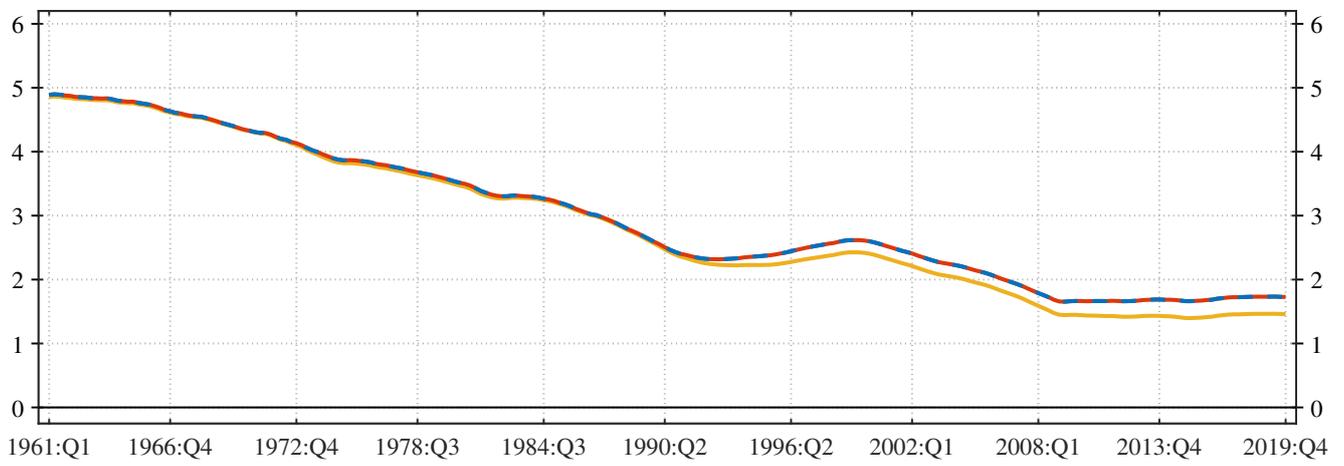
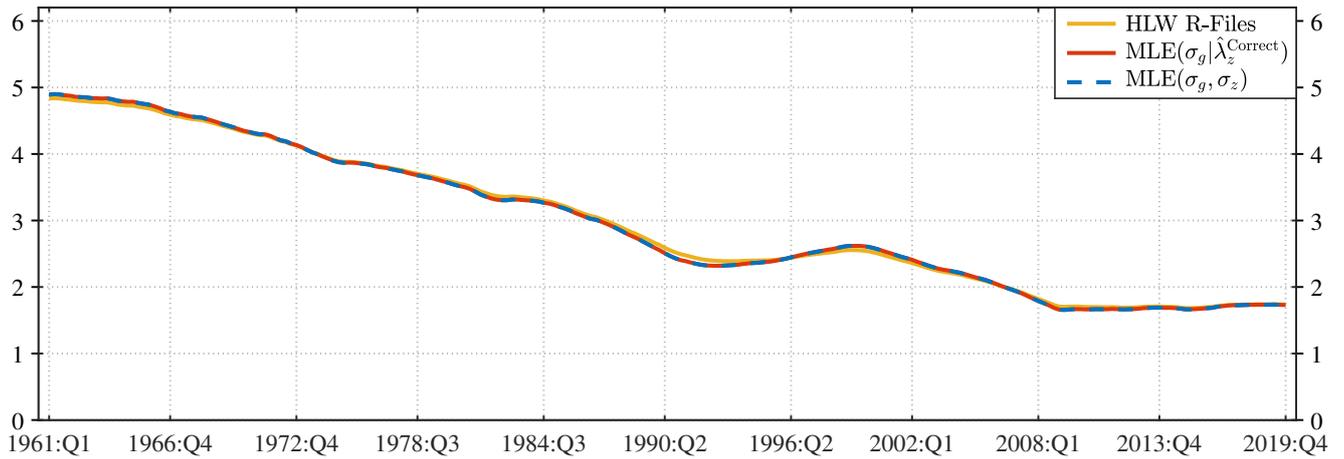
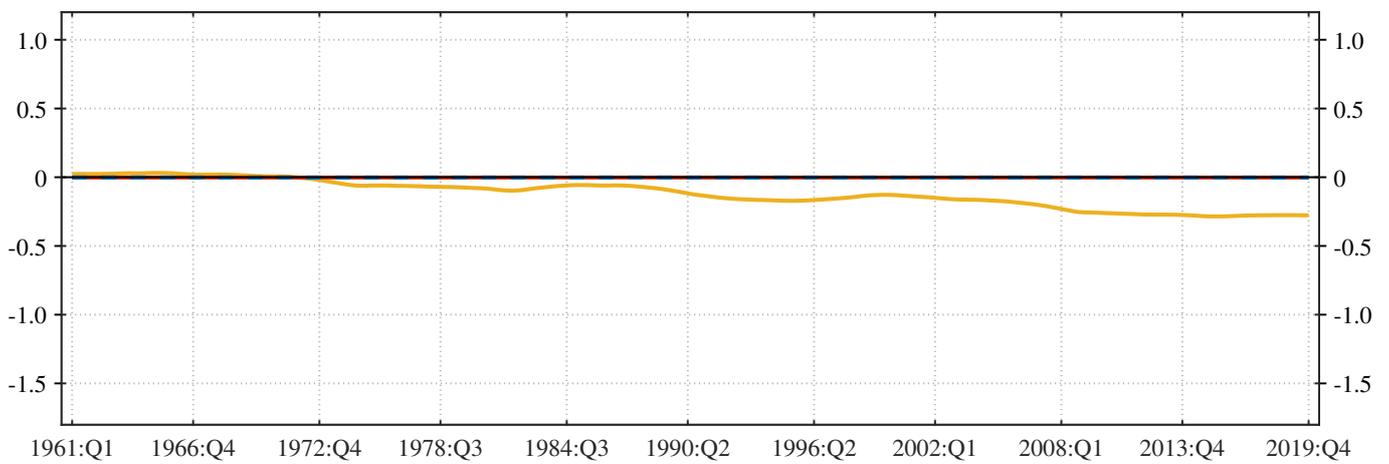
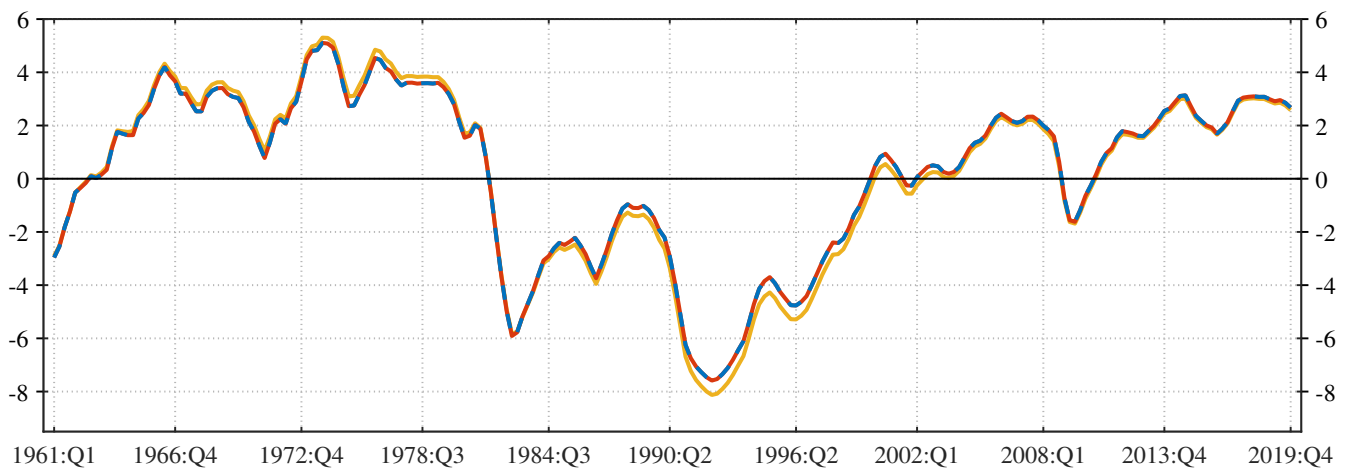
\end{document}